\documentclass[journal,10pt]{IEEEtran}
\pdfoutput=1 
\usepackage{hyperref}
\usepackage{mathtools}
\usepackage{bm}
\usepackage[noadjust]{cite}
\usepackage{xspace}
\usepackage{amsmath}
\usepackage{xcolor}
\usepackage{amsfonts}
\usepackage{setspace}
\usepackage{ifthen}
\usepackage{amsmath} 
 
\usepackage[utf8]{inputenc} 
\usepackage[T1]{fontenc}    
\usepackage{url}            
\usepackage{booktabs}       
\usepackage{amsfonts}       
\usepackage{nicefrac}       
\usepackage{microtype}      
\usepackage{lipsum}
\usepackage{stfloats}

 \newboolean{hidenotes}
\newboolean{draft}
\newboolean{singlecolumn}
\newboolean{icml}
\newboolean{anon}

\newcommand{\TODO}[1]{
  \ifthenelse{\boolean{hidenotes}}
  {}{\textcolor{red}{ TODO: #1 }}
}

\newcommand{\comment}[1]{
  \ifthenelse{\boolean{hidenotes}}
  {}{\textcolor{blue}{ #1 }}
}

\newcommand{\revnum}[2]{
  \ifthenelse{\boolean{hidenotes}}
  {}{\marginpar{\textcolor{red}{ R#1.#2 }}}
}

\renewcommand{\vec}[1]{\bm{#1}}

\newcommand{\norm}[1]{\left\lVert#1\right\rVert}
\newcommand{\nth}[1]{#1^{\mathrm{th}}}

\newcommand{\complex}[1]{\mathbb{C}^{#1}}

\newcommand{\mysection}[1]{\vspace{-.1in}\section{#1}}

\newcommand{\mysubsection}[1]{\vspace{-.05in}\subsection{#1}}

\def\figureautorefname~#1\null{
  Fig.~#1\null
}
\def\sectionautorefname~#1\null{
  Section~#1\null
}
\def\subsectionautorefname~#1\null{
  Subsection~#1\null
}
\def\equationautorefname~#1\null{
  (#1)\null
}

\newcommand{\myfig}[2][placeholder]	
{
  \begin{figure}[htb]
    \centering
    \ifthenelse{\boolean{singlecolumn}}
    {
      \begin{minipage}[c]{.5\textwidth}
	\includegraphics[width=\textwidth]{#1}
      \end{minipage}\hfill
    }
    {\includegraphics[width=\linewidth]{#1}}
    \ifthenelse{\boolean{singlecolumn}}
    {
      \begin{minipage}[c]{.4\textwidth}
      }
      {\vspace{-8mm}}
      \caption{#2}
    \label{fig:#1}
      \ifthenelse{\boolean{singlecolumn}}
      {
      \end{minipage}
    }
    {}
  \end{figure}
}

\newcommand{\myfigfull}[2][placeholder]	
{
  \begin{figure*}[b!]
    \centering
    \includegraphics[width=.8\textwidth]{#1}
    \caption{#2}
    \label{fig:#1}
  \end{figure*}
}

\setboolean{hidenotes}{true}
\setboolean{singlecolumn}{false}
\setboolean{draft}{false}
\setboolean{icml}{false}
\setboolean{anon}{false}

\ifthenelse{\boolean{draft}}
{\usepackage[draft]{graphicx}}
{\usepackage{graphicx}}

\ifthenelse{\boolean{icml}}
{
\usepackage[accepted]{icml2019}
\title{\Large Unsupervised Deep Basis Pursuit:\\Learning inverse problems without ground-truth data}
}
{
\usepackage{geometry}
\geometry{rmargin=1.00in,lmargin=1.00in,top=1in,bottom=1in}
\title{\Large Unsupervised Deep Basis Pursuit:\\Learning inverse problems without ground-truth data}
}

\ifthenelse{\boolean{anon}}
{
\author{
\thanks{$^1$Anonymous institution 1. $^2$ Anonymous Institution 2.$^3$ Anonymous institution 3. $^4$Anonymous Institution 4. . Correspondence to: Anonymous Author <\url{anon@anon.edu}>.
}
Anonymous Authors
\vspace{-.3in}
}
}
{
\author{
\thanks{$^1$Electrical Engineering and Computer Sciences, University of California, Berkeley. $^2$International Computer Science Institute, University of California, Berkeley. $^3$Electrical and Computer Engineering, The University of Texas at Austin. Correspondence to: Jonathan Tamir <\url{jtamir@utexas.edu}>.
This work was first presented at the ISMRM 27th Annual Meeting, Montreal, Canada 2019 \cite{bib:dbp_ismrm}.}
  Jonathan I.\ Tamir$^{1,3}$,\ Stella X.\ Yu$^{1,2}$,\ Michael Lustig$^{1}$
\vspace{-.3in}
}
}

\begin{document}
\maketitle

\begin{abstract}
  Basis pursuit is a compressed sensing optimization in which the $\ell_1$-norm is minimized subject to model error constraints. Here we use a deep neural network prior instead of $\ell_1$-regularization. Using known noise statistics, we jointly learn the prior and reconstruct images without access to ground-truth data. During training, we use alternating minimization across an unrolled iterative network and jointly solve for the neural network weights and training set image reconstructions. At inference, we fix the weights and pass the measurements through the network. We compare reconstruction performance between unsupervised and supervised (i.e. with ground-truth) methods. We hypothesize this technique could be used to learn reconstruction when ground-truth data are unavailable, such as in high-resolution dynamic MRI. \end{abstract}

\mysection{Introduction}
Deep learning in tandem with model-based iterative optimization \cite{bib:lista, Diamond_Unrolled_2017, Schlemper_A, Hammernik_Learning_2017, Aggarwal_MoDL_2018}, i.e.\ model-based deep learning, has shown great promise at solving imaging-based inverse problems beyond the capabilities of compressed sensing \cite{Lustig_Sparse_2007}. These networks typically require hundreds to thousands of examples for training, consisting of pairs of corrupted measurements and the desired ground-truth image. The reconstruction is then trained in an end-to-end fashion, in which data are reconstructed with the network and compared to the ground-truth result. in many cases, collecting a large set of fully sampled data for training is expensive, impractical, or impossible.

In this work, we present an approach to model-based deep learning without access to ground-truth data \cite{Ulyanov_Deep_2018, Lehtinen_Noise2noise_2018, Ong_Low_2018}. We take advantage of (known) noise statistics for each training example and formulate the problem as an extension of basis pursuit denoising \cite{Chen_Atomic_2001} with a deep convolutional neural network (CNN) prior in place of image sparsity. During training, we jointly solve for the CNN weights and the reconstructed training set images. At inference time, we fix the weights and pass the measured data through the network.

As proof of principle, we apply the technique to under-sampled, multi-channel magnetic resonance imaging (MRI). We compare our Deep Basis Pursuit (DBP) formulation with and without supervised learning, as well as to MoDL \cite{Aggarwal_MoDL_2018}, a recently proposed unrolled model-based network that uses ground-truth data for training. We show that in the unsupervised setting, we are able to approach the image reconstruction quality of supervised learning, thus opening the door to applications where collecting fully sampled data is not possible.
 \mysection{Background}
\mysubsection{Imaging Model}
We focus on the discretized linear signal model under additive white Gaussian noise:
\begin{align}
    \vec y = \vec{Ax} + \vec v, \label{eqn:signal_model}
\end{align}
where $\vec x \in \complex{N}$ is the vectorized unknown image, $\vec A \in \complex{M\times N}$ is the discretized forward model describing the imaging system,
$\vec y \in \complex{M}$ is a vector of the acquired measurements, and $\vec v \sim \mathcal{N}_c\left(\vec 0, \sigma^2\vec I\right)$ is a complex-valued Gaussian noise vector.
We are interested in the ill-posed regime, where $M < N$.
To make the inverse problem well-posed, $\vec x$ is commonly solved through a regularized least-squares:
\begin{align}
    \arg\min_{\vec x} \frac{1}{2}\norm{\vec y - \vec{Ax}}_2^2 + \lambda Q(\vec x), \label{eqn:lasso}
\end{align}
where $Q(\vec x)$ is a suitable regularization term, and $\lambda > 0$ is the strength of the regularization.

An alternative, equivalent formulation that directly accounts for the model error due to noise is the constrained problem:
\begin{equation}
\begin{aligned}
& \arg\min_{\vec x}
& & Q(\vec x) \\
& \text{subject to}
& & \norm{\vec y - \vec{Ax}}_2 \leq \epsilon,\label{eqn:bp}
\end{aligned}
\end{equation}
where $\epsilon = \sigma \sqrt{M}$ is the square-root of the expected noise power in the measurements.
When an $\ell_1$-norm is used for regularization, this is known as basis pursuit denoising \cite{Chen_Atomic_2001}, and provides an intuitive formulation as it finds the best (sparsist) representation given a noise error constraint.

\mysubsection{Deep Learning Image Reconstruction}
CNNs have recently been used to solve imaging inverse problems, relying on the network architecture and training data to learn the inverse mapping. When a large corpus of training data is available, it is possible to learn the inverse mapping directly from under-sampled measurements, typically by first transforming the measurements to the image domain either through the adjoint operation $\vec A^*\vec y$ or through a conventional reconstruction.
Except for the initial transformation, these models do not take advantage of knowledge of the imaging system in the network architecture. Thus, they require substantial training data and are prone to overfitting and CNN artifacts \cite{bib:antun2019}.

More recently, network architectures that combine both CNN blocks and data consistency blocks incorporating knowledge of the forward model have grown in popularity, as they allow for robustness against CNN artifacts and training with limited data \cite{Hammernik_Learning_2017, Aggarwal_MoDL_2018}. These architectures are inspired by conventional first-order iterative algorithms intended to solve the unconstrained problem \autoref{eqn:lasso}, and typically alternate between data consistency and manifold projection. To facilitate training with backpropagation, the iterative algorithms are unrolled for a finite number of steps and optimized in an end-to-end manner. As the network is differentiable, gradient updates can be computed through the application of the forward operator with auto-differentiation.

For a particular network architecture, we can view the image reconstruction as a feed-forward network
\begin{align}
    \hat{\vec x}_{\vec w} = \mathcal{F}_{\vec w}\left(\vec y; \vec A\right), \label{eqn:dnn}
\end{align}
where $\mathcal{F}_{\vec w}$ is a deep network parameterized by weights $\vec w$ that operates on the measurements and optionally incorporates knowledge of the forward model. Given a training set of inputs $\{\vec y^{(i)}, \vec A^{(i)}\}_{i=1}^L$ and corresponding ground-truth images $\{\vec x^{(i)}\}_{i=1}^L$, 
the network weights can be trained in a traditional end-to-end fashion by minimizing the average training loss as measured by the loss function $\mathcal{L}$:
\begin{align}
\min_{\vec w} \frac{1}{L}\sum_{i=1}^L \mathcal{L}\left(\hat{\vec x}_{\vec w}^{(i)}, \vec x^{(i)}\right) \label{eqn:loss_sup}.
\end{align}
For inference, the weights are fixed and new measurements are reconstructed through a forward pass of \autoref{eqn:dnn}.

\mysection{Proposed Method}
Inspired by other model-based deep learning architectures \cite{bib:lista, Diamond_Unrolled_2017, Schlemper_A, Aggarwal_MoDL_2018, Hammernik_Learning_2017}, we propose a new unrolled network based on basis pursuit denoising, which we call Deep Basis Pursuit (DBP). 
We assume the noise statistics of the measurements are known and we use them to self-regularize the solution. In turn, we propose to train in an unsupervised fashion in the measurement domain, taking advantage of explicit control of the error between the measurements and the output of the network.
    We first describe the DBP model, and then discuss training the in an unsupervised fashion without ground-truth data.

\mysubsection{Deep Basis Pursuit}
We combine the data consistency constraint of basis pursuit denoising \autoref{eqn:bp} with the $\ell_2$-norm incorporating a CNN auto-encoder.
The DBP optimization is given by
\begin{equation}
\begin{aligned}
& \arg\min_{\vec x}
& & \frac{1}{2}\norm{\mathcal{N}_{\vec w}\left(\vec x)\right)}_2^2  \\
& \text{subject to}
& & \norm{\vec y - \vec{Ax}}_2 \leq \epsilon,\label{eqn:dbp}
\end{aligned}
\end{equation}
where $\mathcal{N}_{\vec{w}}(\vec x) \equiv \vec x - \mathcal{R}_{\vec{w}}(\vec x)$ is a CNN parameterized by weights $\vec w$ that aims to estimate  noise and aliasing \cite{Schlemper_A, Aggarwal_MoDL_2018}. In other words, $\mathcal{R}_{\vec w}(\vec x)$ represents a denoised version of $\vec x$.
In this way, we seek to find the "cleanest" representation of $\vec x$ while allowing for the expected data inconsistency due to noise.

To approximately solve \autoref{eqn:dbp}, we consider an alternating minimization \cite{Aggarwal_MoDL_2018}, repeated $N_1$ times:
\begin{align}
    \vec r_k &= \mathcal{R}_{\vec{w}}(\vec x_{k-1}), \label{eqn:am1}\\
    \vec x_k &= \arg\min_{\vec{x}} \frac{1}{2}\norm{\vec x - \vec r_k}_2^2 \ \text{s.t.} \  \norm{\vec y - \vec A \vec x}_2 \leq \epsilon. \label{eqn:am2}
\end{align}
Subproblem \autoref{eqn:am1} is a forward pass through the CNN. Subproblem \autoref{eqn:am2} is convex and can solved with ADMM \cite{Boyd_Distributed_2011}. We introduce the slack variable $\vec z = \vec{Ax}$ and the dual variable $\vec u$, and apply the following update steps, repeated $N_2$ times:
\begin{align}
    \vec x_l &= \left(\rho\vec{A}^*\vec{A}+\vec I\right)^{-1}\left( \rho\vec A^*\left(\vec z_{l-1} - \vec u_{l-1}\right) + \vec r_k\right), \label{eqn:admm1}\\
    \vec z_l &= \vec y + \mathrm{L2Proj}\left(\vec{Ax}_l+\vec u_{l-1} - \vec y, \epsilon\right), \label{eqn:admm2}\\
    \vec u_l &= \vec u_{l-1} + \vec{Ax}_l + \vec z_l, \label{eqn:admm3}
\end{align}
where $\rho > 0$ is the ADMM penalty parameter and $\mathrm{L2Proj}(\vec z, \epsilon)$ is the projection of $\vec z$ onto the $\ell_2$-ball of radius $\epsilon$.
The update steps are amenable to matrix-free optimization, as the forward and adjoint calculations can be represented as computationally efficient operators. In particular, subproblem \autoref{eqn:admm1} can be approximately solved with $N_3$ iterations of the Conjugate Gradient Method.

Altogether, we can view DBP as an unrolled optimization alternating between CNN layers and data consistency layers, as shown in \autoref{fig:unrolled_network}. At each layer, the same CNN is used, though it is possible in general to relax this requirement \cite{Schlemper_A}.
For a fixed CNN $\mathcal{R}_{\vec w}$, the DBP model is a special case of \autoref{eqn:dnn}:
$\hat{\vec x}_{\tilde{\vec w}} \equiv \mathcal{F}_{\tilde{\vec w}}(\vec y;  \vec A, \epsilon)$,
where $\tilde{\vec w} = (\vec w, \rho)$ are the network parameters, and the network uses measurements together with knowledge of the system and noise power to return an estimate of the image.
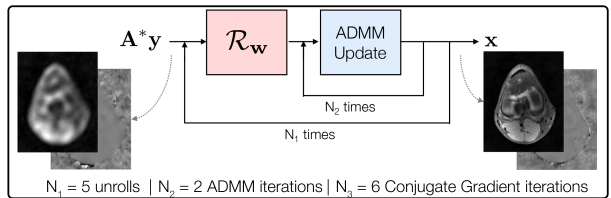
\myfig[unrolled_network]{DBP network architecture, consisting of $N_1$ unrolled iterations alternating between CNN layers and data consistency layers. The data consistency layer is solved through $N_2$ unrolled ADMM iterations.}

\mysubsection{Unsupervised Learning}
When both input $\{\vec y^{(i)}, \vec A^{(i)}, \epsilon^{(i)}\}_{i=1}^L$ and ground-truth $\{\vec x^{(i)}\}_{i=1}^L$ training data are
available, the network weights can be trained in a traditional end-to-end fashion according to \autoref{eqn:loss_sup}.
When ground-truth data are not available, we consider a loss function $\hat{\mathcal{L}}$ imposed in the measurement domain:
\begin{align}
\min_{\tilde{\vec w}} \frac{1}{L}\sum_{i=1}^L \hat{\mathcal{L}} \left(\vec{A}^{(i)}\hat{\vec x}_{\tilde{\vec w}}^{(i)}, \vec y^{(i)}\right) \label{eqn:loss_unsup}.
\end{align}
The measurement loss can be a useful surrogate for the true loss, as the measurements contain (noisy) information about the ground-truth image \cite{Lehtinen_Noise2noise_2018, bib:batson_noise2self_2019}. Thus, we may hope to learn about the image statistics given a large-enough training set that includes a diversity of measurements. 
 \mysection{Methods}
We consider the application to under-sampled, multi-channel MRI. The MRI reconstruction task is well-suited to DBP, as the noise statistics are Gaussian and can be measured during a short pre-scan.
We first describe the multi-channel MRI forward operator and general sampling strategy. Then we discuss the experimental setup, including the dataset and implementation details.

\mysubsection{Multi-channel MRI forward operator}\label{subsec:mri_operator}
In multi-channel MRI, the signal is measured by an array of receive coils distributed around an object, each with a spatially-varying sensitivity profile. In the measurement model, the image is linearly mixed with each coil sensitivity profile, Fourier transformed, and sampled. We can describe the measurement model as
$\vec A = \begin{bmatrix}(\vec{PFS}_1)^\top & \cdots & (\vec{PFS}_C)^\top\end{bmatrix}^\top \in \complex{M\times N}$,
where $C$ is the number of receive coils, $\vec S_c \in \complex{N\times N}$ is a diagonal operator containing the spatial sensitivity profile of the $\nth{c}$ coil along the diagonal, $\vec F$ is the Fourier transform operator, and $\vec P \in \{0, 1\}^{\frac{M}{C}\times N}$ is a diagonal operator that selects the sampled frequencies.

\mysubsection{Experimental setup}

\subsubsection*{Data}
We used the "Stanford Fully Sampled 3D FSE Knees" dataset from \url{mridata.org}, containing 3D Cartesian proton-density knee scans of 20 healthy volunteers. Each 3D volume consisted of 320 slices with matrix size $320 \times 256$ and was scanned with an 8-channel receive coil array. Although each slice is fully sampled, in practice the ``ground-truth'' data itself has noise.
To aid in experimental comparison, ``noise-free'' ground-truth data were created by averaging the data from seven adjacent slices. 
For each slice, the spatial sensitivity profiles of each coil were estimated using ESPIRiT \cite{bib:uecker2014}, a self-calibrated parallel imaging method. Ground-truth images were reconstructed by solving \autoref{eqn:lasso} using the fully sampled data without regularization.
Each slice was then passed through the forward model and retrospectively under-sampled using a different variable-density Poisson-disc sampling pattern \cite{Lustig_Sparse_2007, Uecker_bart_2018} with a $16\times 16$ calibration region and acceleration factor $R\approx 12$. 
Slices from the first 16 volunteers were used for training, discarding the first and last 20 edge slices of each volume (4,384 slices). Similarly, slices from the next two volunteers were used for validation (548 slices), and slices from the last two volunteers were used for testing (548 slices).
We added complex-valued Gaussian noise with standard deviation $\sigma=0.01$ to the noise-free, averaged data.

\subsubsection*{Implementation}
For all experiments we used a Euclidean norm loss function for training. When training with ground-truth (supervised), the loss was applied in the image domain. For unsupervised training, the loss was applied in the measurement (Fourier) domain.
We used a U-Net architecture \cite{bib:ronnebergerUNet} for the CNN auto-encoder, with separate input channels for the real and imaginary components.
The U-Net consisted of three encoding layers with ReLU activation functions and 64, 128, and 256 channels, respectively, followed by three similar decoding layers. A final convolutional layer with no activation function mapped the decoder back to two channels. All convolutions used a $3\times 3$ kernel size.
For comparison, MoDL \cite{Aggarwal_MoDL_2018} was also implemented using the same unrolled parameters and CNN architecture.
All networks were implemented in PyTorch.

\subsubsection*{Evaluation}
DBP was separately trained with and without ground-truth data. We also trained MoDL with ground-truth data.
In addition, we also evaluated parallel imaging and compressed sensing (PICS) \cite{Lustig_Sparse_2007} using BART \cite{Uecker_bart_2018}, with $\ell_1$-Wavelet regularization parameter optimized over the validation set.
Normalized root mean-squared error (NRMSE) was used to compare reconstructions.

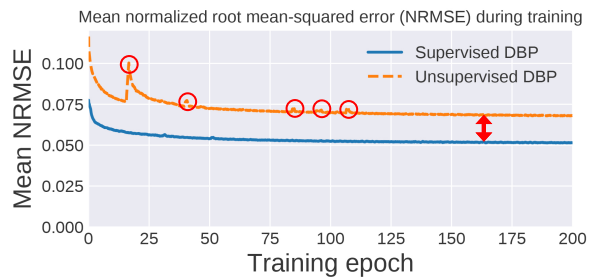
 \myfig[training_loss]{Mean NRMSE on training set for supervised and unsupervised DBP. Unsupervised DBP has a performance drop (red double-arrow), and noisier updates (red circles).}
\mysection{Results}
\autoref{fig:training_loss} shows the mean NRMSE on the training set for each epoch. In addition to a performance gap between supervised and unsupervised learning, unsupervised DBP has noisier updates, likely because the loss function in the measurement domain is a noisy surrogate to the NRMSE. \autoref{fig:nrmse_unrolls} shows the NRMSE across the validation set for different numbers of unrolls during inference. Even though the networks were trained with 5 unrolls, best performance is seen for different number of unrolls (6, 10 and 12 for MoDL, unsupervised DBP, and supervised DBP, respectively). Compared to MoDL, the DBP formulation behaves more stably as the number of unrolls increases, which may be due to the hard data consistency constraint.
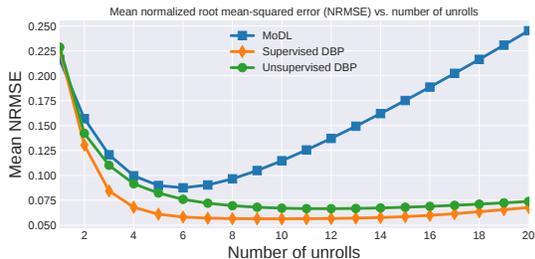
\myfig[nrmse_unrolls]{Mean NRMSE vs. number of rolls at inference time on validation set. All networks were trained with 5 unrolls.}

\autoref{fig:nrmse_box_plot} shows a box plot of the NRMSE on the test set for the different networks and for PICS. Both 5 unrolls and the validation-optimized number of unrolls are shown. At the optimal number of unrolls, unsupervised DBP outperforms PICS.
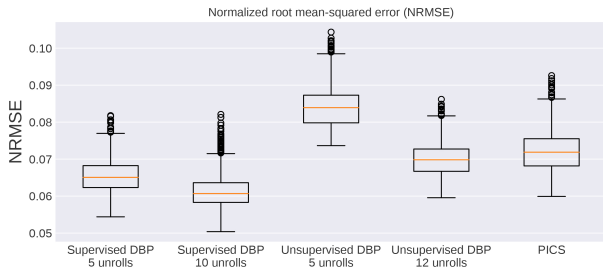
\myfig[nrmse_box_plot]{Box plot of test set NRMSE for supervised and unsupervised DBP at two different unrolls -- the first matching unrolls at training, and the second chosen to minimize validation set mean NRMSE. Also shown is PICS NRMSE for optimized regularization on validation set.}

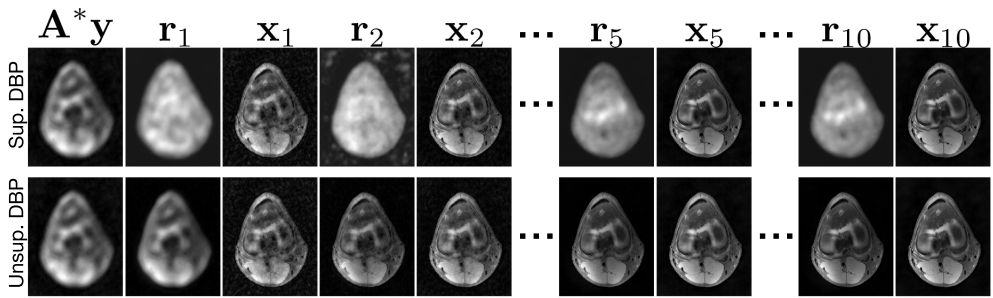
\myfigfull[layer_vis]{Output images after different layers in the unrolled network: $\vec r_k$ and $\vec x_k$ are the $\nth{k}$ application of the CNN and the data consistency updates, respectively. The supervised CNN amplifies and denoises features to a greater effect compared to the unsupervised CNN.}
\autoref{fig:layer_vis} shows some of the intermediate output stages for the supervised and unsupervised DBP networks, indicating that similar structure is learned in both CNNS; however, the supervised DBP appears to better amplify and denoise features in the image.
The magnitude reconstructions and error maps of a representative slice from the test set are shown in \autoref{fig:test_img_compare}. Supervised DBP achieves the lowest error, followed by unsupervised DBP, PICS, and MoDL. Small details in edge sharpness are retained with DBP, but reduced with MoDL and PICS.

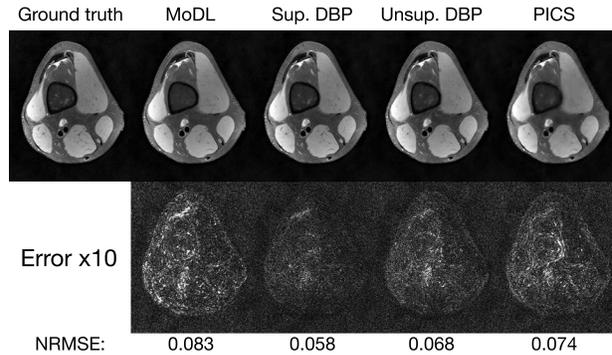
\myfig[test_img_compare]{Comparison of different reconstruction methods to ground-truth image from the test set.} \mysection{Discussion and Conclusion}
There are strong connections to iterative optimization and unrolled deep networks \cite{Ong_Low_2018,Papyan_Convolutional_2017,Sulam_Multilayer_2018}. Jointly optimizing over the images and weights could be viewed a non-linear extension to dictionary learning. Nonetheless, there is a cost in reconstruction error when moving to unsupervised learning, highlighting the importance of a large training data set to offset the missing ground-truth information. The choice of measurement loss function and data SNR may also greatly impact the quality. Fortunately, in many practical settings there is an abundance of under-sampled or corrupted measurement data available for training.

In conclusion, the combination of basis pursuit denoising and deep learning can take advantage of under-sampled data and provide a means to learn model-based deep learning reconstructions without access to ground-truth images. 

\newpage
\onecolumn{
\section*{Acknowledgements}
This work received research support from GE Healthcare and from National Institutes of Health (NIH) grants NIH R01EB026136, NIH R01EB009690, NIH R01HL136965, and NIH U24EB029240.
 \bibliographystyle{IEEEtran}  
\bibliography{refs}
}

\end{document}